\documentclass[final]{svjour3_compact_authors}
\usepackage{graphicx}
\usepackage{rotating}
\usepackage{amssymb}
\usepackage{mathptmx}
\usepackage{siunitx}
\usepackage{amsmath}

\usepackage[
    backend=bibtex,
    natbib=true,
    url=false, 
    doi=false,
    eprint=true,
    giveninits=true,
    maxnames=3,
    sorting=none
]{biblatex}
\AtEveryBibitem{\clearfield{title}}
\renewbibmacro{in:}{}
\DefineBibliographyStrings{english}{%
  page             = {\ifbibliography{}{\adddot}},
  pages            = {\ifbibliography{}{\adddot}},
}
\DeclareFieldFormat{postnote}{#1}

\makeatletter
\def\blx@bblfile@biber{%
  \blx@secinit
  \begingroup
  \blx@bblstart
\begingroup
\makeatletter
\@ifundefined{ver@biblatex.sty}
  {\@latex@error
     {Missing 'biblatex' package}
     {The bibliography requires the 'biblatex' package.}
      \aftergroup }
  {}
\endgroup

\sortlist[entry]{none/global/}
  \entry{Henning:2017nuy}{article}{}
    \true{moreauthor}
    \name{author}{1}{}{%
      {{hash=HJW}{%
         family={Henning},
         familyi={H\bibinitperiod},
         given={J.\bibnamedelima W.},
         giveni={J\bibinitperiod\bibinitdelim W\bibinitperiod},
      }}%
    }
    \strng{namehash}{HJW+1}
    \strng{fullhash}{HJW+1}
    \field{labelnamesource}{author}
    \field{labeltitlesource}{title}
    \verb{doi}
    \verb 10.3847/1538-4357/aa9ff4
    \endverb
    \field{number}{2}
    \field{pages}{97}
    \field{title}{{Measurements of the Temperature and E-Mode Polarization of
  the CMB from 500 Square Degrees of SPTpol Data}}
    \field{volume}{852}
    \field{journaltitle}{Astrophys. J.}
    \field{eprinttype}{arXiv}
    \field{eprintclass}{astro-ph.CO}
    \field{year}{2018}
  \endentry

  \entry{Wu:2019hek}{article}{}
    \true{moreauthor}
    \name{author}{1}{}{%
      {{hash=WWLK}{%
         family={Wu},
         familyi={W\bibinitperiod},
         given={W.\bibnamedelima L.\bibnamedelima K.},
         giveni={W\bibinitperiod\bibinitdelim L\bibinitperiod\bibinitdelim
  K\bibinitperiod},
      }}%
    }
    \strng{namehash}{WWLK+1}
    \strng{fullhash}{WWLK+1}
    \field{labelnamesource}{author}
    \field{labeltitlesource}{title}
    \verb{eprint}
    \verb 1905.05777
    \endverb
    \field{title}{{A Measurement of the Cosmic Microwave Background Lensing
  Potential and Power Spectrum from 500 deg$^2$ of SPTpol Temperature and
  Polarization Data}}
    \field{eprinttype}{arXiv}
    \field{eprintclass}{astro-ph.CO}
    \field{year}{2019}
  \endentry

  \entry{Sherwin:2016tyf}{article}{}
    \true{moreauthor}
    \name{author}{1}{}{%
      {{hash=SBD}{%
         family={Sherwin},
         familyi={S\bibinitperiod},
         given={Blake\bibnamedelima D.},
         giveni={B\bibinitperiod\bibinitdelim D\bibinitperiod},
      }}%
    }
    \strng{namehash}{SBD+1}
    \strng{fullhash}{SBD+1}
    \field{labelnamesource}{author}
    \field{labeltitlesource}{title}
    \verb{doi}
    \verb 10.1103/PhysRevD.95.123529
    \endverb
    \field{number}{12}
    \field{pages}{123529}
    \field{title}{{Two-season Atacama Cosmology Telescope polarimeter lensing
  power spectrum}}
    \field{volume}{D95}
    \field{journaltitle}{Phys. Rev.}
    \field{eprinttype}{arXiv}
    \field{eprintclass}{astro-ph.CO}
    \field{year}{2017}
  \endentry

  \entry{Manzotti:2017net}{article}{}
    \true{moreauthor}
    \name{author}{1}{}{%
      {{hash=MA}{%
         family={Manzotti},
         familyi={M\bibinitperiod},
         given={A.},
         giveni={A\bibinitperiod},
      }}%
    }
    \strng{namehash}{MA+1}
    \strng{fullhash}{MA+1}
    \field{labelnamesource}{author}
    \field{labeltitlesource}{title}
    \verb{doi}
    \verb 10.3847/1538-4357/aa82bb
    \endverb
    \field{pages}{1}
    \field{title}{{CMB Polarization B-mode Delensing with SPTpol and Herschel}}
    \field{volume}{846}
    \field{journaltitle}{Astrophys. J.}
    \field{eprinttype}{arXiv}
    \field{eprintclass}{astro-ph.CO}
    \field{year}{2017}
  \endentry

  \entry{Ade:2018gkx}{article}{}
    \true{moreauthor}
    \name{author}{1}{}{%
      {{hash=APAR}{%
         family={Ade},
         familyi={A\bibinitperiod},
         given={P.\bibnamedelima A.\bibnamedelima R.},
         giveni={P\bibinitperiod\bibinitdelim A\bibinitperiod\bibinitdelim
  R\bibinitperiod},
      }}%
    }
    \strng{namehash}{APAR+1}
    \strng{fullhash}{APAR+1}
    \field{labelnamesource}{author}
    \field{labeltitlesource}{title}
    \verb{doi}
    \verb 10.1103/PhysRevLett.121.221301
    \endverb
    \field{pages}{221301}
    \field{title}{{BICEP2 / Keck Array x: Constraints on Primordial
  Gravitational Waves using Planck, WMAP, and New BICEP2/Keck Observations
  through the 2015 Season}}
    \field{volume}{121}
    \field{journaltitle}{Phys. Rev. Lett.}
    \field{eprinttype}{arXiv}
    \field{eprintclass}{astro-ph.CO}
    \field{year}{2018}
  \endentry

  \entry{Bleem:2014iim}{article}{}
    \true{moreauthor}
    \name{author}{1}{}{%
      {{hash=BLE}{%
         family={Bleem},
         familyi={B\bibinitperiod},
         given={L.\bibnamedelima E.},
         giveni={L\bibinitperiod\bibinitdelim E\bibinitperiod},
      }}%
    }
    \strng{namehash}{BLE+1}
    \strng{fullhash}{BLE+1}
    \field{labelnamesource}{author}
    \field{labeltitlesource}{title}
    \verb{doi}
    \verb 10.1088/0067-0049/216/2/27
    \endverb
    \field{number}{2}
    \field{pages}{27}
    \field{title}{{Galaxy Clusters Discovered via the Sunyaev-Zel'dovich Effect
  in the 2500-square-degree SPT-SZ survey}}
    \field{volume}{216}
    \field{journaltitle}{Astrophys. J. Suppl.}
    \field{eprinttype}{arXiv}
    \field{eprintclass}{astro-ph.CO}
    \field{year}{2015}
  \endentry

  \entry{Hilton:2017gal}{article}{}
    \true{moreauthor}
    \name{author}{1}{}{%
      {{hash=HM}{%
         family={Hilton},
         familyi={H\bibinitperiod},
         given={Matt},
         giveni={M\bibinitperiod},
      }}%
    }
    \strng{namehash}{HM+1}
    \strng{fullhash}{HM+1}
    \field{labelnamesource}{author}
    \field{labeltitlesource}{title}
    \verb{doi}
    \verb 10.3847/1538-4365/aaa6cb
    \endverb
    \field{number}{1}
    \field{pages}{20}
    \field{title}{{The Atacama Cosmology Telescope: The Two-Season ACTPol
  Sunyaev-Zel'dovich Effect Selected Cluster Catalog}}
    \field{volume}{235}
    \field{journaltitle}{Astrophys. J. Suppl.}
    \field{eprinttype}{arXiv}
    \field{eprintclass}{astro-ph.CO}
    \field{year}{2018}
  \endentry

  \entry{Vieira:2009ru}{article}{}
    \true{moreauthor}
    \name{author}{1}{}{%
      {{hash=VJD}{%
         family={Vieira},
         familyi={V\bibinitperiod},
         given={J.\bibnamedelima D.},
         giveni={J\bibinitperiod\bibinitdelim D\bibinitperiod},
      }}%
    }
    \strng{namehash}{VJD+1}
    \strng{fullhash}{VJD+1}
    \field{labelnamesource}{author}
    \field{labeltitlesource}{title}
    \verb{doi}
    \verb 10.1088/0004-637X/719/1/763
    \endverb
    \field{pages}{763\bibrangedash 783}
    \field{title}{{Extragalactic millimeter-wave sources in South Pole
  Telescope survey data: source counts, catalog, and statistics for an 87
  square-degree field}}
    \field{volume}{719}
    \field{journaltitle}{Astrophys. J.}
    \field{eprinttype}{arXiv}
    \field{eprintclass}{astro-ph.CO}
    \field{year}{2010}
  \endentry

  \entry{Gralla:2019jyc}{article}{}
    \true{moreauthor}
    \name{author}{1}{}{%
      {{hash=GMB}{%
         family={Gralla},
         familyi={G\bibinitperiod},
         given={Megan\bibnamedelima B.},
         giveni={M\bibinitperiod\bibinitdelim B\bibinitperiod},
      }}%
    }
    \strng{namehash}{GMB+1}
    \strng{fullhash}{GMB+1}
    \field{labelnamesource}{author}
    \field{labeltitlesource}{title}
    \verb{eprint}
    \verb 1905.04592
    \endverb
    \field{title}{{Atacama Cosmology Telescope: Dusty star-forming galaxies and
  active galactic nuclei in the equatorial survey}}
    \field{eprinttype}{arXiv}
    \field{eprintclass}{astro-ph.GA}
    \field{year}{2019}
  \endentry

  \entry{Whitehorn:2016njg}{article}{}
    \true{moreauthor}
    \name{author}{1}{}{%
      {{hash=WN}{%
         family={Whitehorn},
         familyi={W\bibinitperiod},
         given={N.},
         giveni={N\bibinitperiod},
      }}%
    }
    \strng{namehash}{WN+1}
    \strng{fullhash}{WN+1}
    \field{labelnamesource}{author}
    \field{labeltitlesource}{title}
    \verb{doi}
    \verb 10.3847/0004-637X/830/2/143
    \endverb
    \field{number}{2}
    \field{pages}{143}
    \field{title}{{Millimeter Transient Point Sources in the SPTpol 100 Square
  Degree Survey}}
    \field{volume}{830}
    \field{journaltitle}{Astrophys. J.}
    \field{eprinttype}{arXiv}
    \field{eprintclass}{astro-ph.HE}
    \field{year}{2016}
  \endentry

  \entry{Benson:2014qhw}{article}{}
    \true{moreauthor}
    \name{author}{1}{}{%
      {{hash=BBA}{%
         family={Benson},
         familyi={B\bibinitperiod},
         given={B.\bibnamedelima A.},
         giveni={B\bibinitperiod\bibinitdelim A\bibinitperiod},
      }}%
    }
    \strng{namehash}{BBA+1}
    \strng{fullhash}{BBA+1}
    \field{labelnamesource}{author}
    \field{labeltitlesource}{title}
    \field{booktitle}{{Proceedings, SPIE Astronomical Telescopes +
  Instrumentation 2014: Millimeter, Submillimeter, and Far-Infrared Detectors
  and Instrumentation for Astronomy VII: Montreal, Quebec, Canada, June 24-27,
  2014}}
    \verb{doi}
    \verb 10.1117/12.2057305
    \endverb
    \field{pages}{91531P}
    \field{title}{{SPT-3G: A Next-Generation Cosmic Microwave Background
  Polarization Experiment on the South Pole Telescope}}
    \field{volume}{9153}
    \field{journaltitle}{Proc. SPIE Int. Soc. Opt. Eng.}
    \field{eprinttype}{arXiv}
    \field{eprintclass}{astro-ph.IM}
    \field{year}{2014}
  \endentry

  \entry{Bender:2018dwl}{article}{}
    \true{moreauthor}
    \name{author}{1}{}{%
      {{hash=BAN}{%
         family={Bender},
         familyi={B\bibinitperiod},
         given={Amy\bibnamedelima N.},
         giveni={A\bibinitperiod\bibinitdelim N\bibinitperiod},
      }}%
    }
    \strng{namehash}{BAN+2}
    \strng{fullhash}{BAN+2}
    \field{labelnamesource}{author}
    \field{labeltitlesource}{title}
    \field{booktitle}{{Proceedings, SPIE Astronomical Telescopes +
  Instrumentation 2018: Modeling, Systems Engineering, and Project Management
  for Astronomy VIII: Austin, USA, June 10-15, 2018}}
    \verb{doi}
    \verb 10.1117/12.2312426
    \endverb
    \field{pages}{1070803}
    \field{title}{{Year 2 instrument status from the SPT-3G cosmic microwave
  background receiver (Conference Presentation)}}
    \field{volume}{10708}
    \field{journaltitle}{Proc. SPIE Int. Soc. Opt. Eng.}
    \field{eprinttype}{arXiv}
    \field{eprintclass}{astro-ph.IM}
    \field{year}{2018}
  \endentry

  \entry{Carter:2018unr}{article}{}
    \true{moreauthor}
    \name{author}{1}{}{%
      {{hash=CFW}{%
         family={Carter},
         familyi={C\bibinitperiod},
         given={F.\bibnamedelima W.},
         giveni={F\bibinitperiod\bibinitdelim W\bibinitperiod},
      }}%
    }
    \strng{namehash}{CFW+1}
    \strng{fullhash}{CFW+1}
    \field{labelnamesource}{author}
    \field{labeltitlesource}{title}
    \verb{doi}
    \verb 10.1007/s10909-018-1910-7
    \endverb
    \field{number}{5-6}
    \field{pages}{695\bibrangedash 702}
    \field{title}{{Tuning SPT-3G Transition-Edge-Sensor Electrical Properties
  with a Four-Layer Ti--Au--Ti--Au Thin-Film Stack}}
    \field{volume}{193}
    \field{journaltitle}{J. Low. Temp. Phys.}
    \field{year}{2018}
  \endentry

  \entry{Henning:2012fz}{article}{}
    \true{moreauthor}
    \name{author}{1}{}{%
      {{hash=HJW}{%
         family={Henning},
         familyi={H\bibinitperiod},
         given={J.\bibnamedelima W.},
         giveni={J\bibinitperiod\bibinitdelim W\bibinitperiod},
      }}%
    }
    \strng{namehash}{HJW+1}
    \strng{fullhash}{HJW+1}
    \field{labelnamesource}{author}
    \field{labeltitlesource}{title}
    \verb{doi}
    \verb 10.1117/12.927172
    \endverb
    \field{pages}{84523A}
    \field{title}{{Feedhorn-coupled TES polarimeter camera modules at 150 GHz
  for CMB polarization measurements with SPTpol}}
    \field{volume}{8452}
    \field{journaltitle}{Proc. SPIE Int. Soc. Opt. Eng.}
    \field{eprinttype}{arXiv}
    \field{eprintclass}{astro-ph.IM}
    \field{year}{2012}
  \endentry

  \entry{Li:2016}{article}{}
    \true{moreauthor}
    \name{author}{1}{}{%
      {{hash=LD}{%
         family={Li},
         familyi={L\bibinitperiod},
         given={D.},
         giveni={D\bibinitperiod},
      }}%
    }
    \strng{namehash}{LD+1}
    \strng{fullhash}{LD+1}
    \field{labelnamesource}{author}
    \field{labeltitlesource}{title}
    \field{number}{1-2}
    \field{pages}{66\bibrangedash 73}
    \field{title}{AlMn Transition Edge Sensors for Advanced ACTPol}
    \field{volume}{184}
    \field{journaltitle}{J. Low. Temp. Phys.}
    \field{year}{2016}
  \endentry

  \entry{Abazajian:2016yjj}{article}{}
    \true{moreauthor}
    \name{author}{1}{}{%
      {{hash=AKN}{%
         family={Abazajian},
         familyi={A\bibinitperiod},
         given={Kevork\bibnamedelima N.},
         giveni={K\bibinitperiod\bibinitdelim N\bibinitperiod},
      }}%
    }
    \strng{namehash}{AKN+1}
    \strng{fullhash}{AKN+1}
    \field{labelnamesource}{author}
    \field{labeltitlesource}{title}
    \verb{eprint}
    \verb 1610.02743
    \endverb
    \field{title}{{CMB-S4 Science Book, First Edition}}
    \field{eprinttype}{arXiv}
    \field{eprintclass}{astro-ph.CO}
    \field{year}{2016}
  \endentry

  \entry{Posada:2016}{article}{}
    \true{moreauthor}
    \name{author}{1}{}{%
      {{hash=PCM}{%
         family={Posada},
         familyi={P\bibinitperiod},
         given={Chrystian\bibnamedelima M.},
         giveni={C\bibinitperiod\bibinitdelim M\bibinitperiod},
      }}%
    }
    \strng{namehash}{PCM+1}
    \strng{fullhash}{PCM+1}
    \field{labelnamesource}{author}
    \field{labeltitlesource}{title}
    \field{pages}{9914}
    \field{title}{{Large arrays of dual-polarized multichroic TES detectors for
  CMB measurements with the SPT-3G receiver }}
    \field{volume}{9914}
    \field{journaltitle}{Proc. SPIE Int. Soc. Opt. Eng.}
    \field{year}{2016}
  \endentry

  \entry{Bender:2019}{article}{}
    \true{moreauthor}
    \name{author}{1}{}{%
      {{hash=BAN}{%
         family={Bender},
         familyi={B\bibinitperiod},
         given={A.\bibnamedelima N},
         giveni={A\bibinitperiod\bibinitdelim N},
      }}%
    }
    \strng{namehash}{BAN+1}
    \strng{fullhash}{BAN+1}
    \field{labelnamesource}{author}
    \field{labeltitlesource}{title}
    \field{title}{On-sky performance of the SPT-3G frequency domain On-sky
  performance of the SPT-3G frequency domain multiplexed readout}
    \field{journaltitle}{J. Low Temp. Phys.}
    \field{year}{2019}
  \endentry

  \entry{Dutcher:2018fod}{article}{}
    \true{moreauthor}
    \name{author}{1}{}{%
      {{hash=DD}{%
         family={Dutcher},
         familyi={D\bibinitperiod},
         given={D.},
         giveni={D\bibinitperiod},
      }}%
    }
    \strng{namehash}{DD+1}
    \strng{fullhash}{DD+1}
    \field{labelnamesource}{author}
    \field{labeltitlesource}{title}
    \verb{doi}
    \verb 10.1117/12.2312451
    \endverb
    \field{pages}{107081Z}
    \field{title}{{Characterization and performance of the second-year SPT-3G
  focal plane}}
    \field{volume}{10708}
    \field{journaltitle}{Proc. SPIE Int. Soc. Opt. Eng.}
    \field{eprinttype}{arXiv}
    \field{eprintclass}{astro-ph.IM}
    \field{year}{2018}
  \endentry

  \entry{Nadolski:2018hck}{article}{}
    \true{moreauthor}
    \name{author}{1}{}{%
      {{hash=NA}{%
         family={Nadolski},
         familyi={N\bibinitperiod},
         given={A.},
         giveni={A\bibinitperiod},
      }}%
    }
    \strng{namehash}{NA+1}
    \strng{fullhash}{NA+1}
    \field{labelnamesource}{author}
    \field{labeltitlesource}{title}
    \field{booktitle}{{Proceedings, SPIE Astronomical Telescopes +
  Instrumentation 2018: Modeling, Systems Engineering, and Project Management
  for Astronomy VIII: Austin, USA, June 10-15, 2018}}
    \verb{doi}
    \verb 10.1117/12.2315674
    \endverb
    \field{pages}{1070843}
    \field{title}{{Broadband anti-reflective coatings for cosmic microwave
  background experiments}}
    \field{volume}{10708}
    \field{journaltitle}{Proc. SPIE Int. Soc. Opt. Eng.}
    \field{eprinttype}{arXiv}
    \field{eprintclass}{astro-ph.IM}
    \field{year}{2018}
  \endentry
\endsortlist

  \blx@bblend
  \endgroup
  \csnumgdef{blx@labelnumber@\the\c@refsection}{0}}
\makeatother

\journalname{Journal of Low Temperature Physics}

\begin{document}

\newcommand{\hdblarrow}{H\makebox[0.9ex][l]{$\downdownarrows$}-}
\title{Performance of Al-Mn Transition-Edge Sensor Bolometers in SPT-3G}

\author{
A.~J.~Anderson\textsuperscript{1,2} \and 
P.~A.~R.~Ade\textsuperscript{3} \and 
Z.~Ahmed\textsuperscript{4,5} \and 
J.~S.~Avva\textsuperscript{6} \and 
P.~S.~Barry\textsuperscript{7,2} \and 
R.~Basu Thakur\textsuperscript{2} \and 
A.~N.~Bender\textsuperscript{7,2} \and 
B.~A.~Benson\textsuperscript{1,2,8} \and 
L.~Bryant\textsuperscript{9} \and 
K.~Byrum\textsuperscript{7} \and 
J.~E.~Carlstrom\textsuperscript{2,9,10,7,8} \and 
F.~W.~Carter\textsuperscript{7,2} \and 
T.~W.~Cecil\textsuperscript{7} \and 
C.~L.~Chang\textsuperscript{7,2,8} \and 
H.-M.~Cho\textsuperscript{5} \and 
J.~F.~Cliche\textsuperscript{11} \and 
A.~Cukierman\textsuperscript{6} \and 
T.~de~Haan\textsuperscript{6} \and 
E.~V.~Denison\textsuperscript{12} \and 
J.~Ding\textsuperscript{13} \and 
M.~A.~Dobbs\textsuperscript{11,14} \and 
D.~Dutcher\textsuperscript{2,10} \and 
W.~Everett\textsuperscript{15} \and 
K.~R.~Ferguson\textsuperscript{16} \and 
A.~Foster\textsuperscript{17} \and 
J.~Fu\textsuperscript{18} \and 
J.~Gallicchio\textsuperscript{2,19} \and 
A.~E.~Gambrel\textsuperscript{2} \and 
R.~W.~Gardner\textsuperscript{9} \and 
A.~Gilbert\textsuperscript{11} \and 
J.~C.~Groh\textsuperscript{6} \and 
S.~T.~Guns\textsuperscript{6} \and 
R.~Guyser\textsuperscript{18} \and 
N.~W.~Halverson\textsuperscript{15,20} \and 
A.~H.~Harke-Hosemann\textsuperscript{7,18} \and 
N.~L.~Harrington\textsuperscript{6} \and 
J.~W.~Henning\textsuperscript{7,2} \and 
G.~C.~Hilton\textsuperscript{12} \and 
W.~L.~Holzapfel\textsuperscript{6} \and 
D.~Howe\textsuperscript{21} \and 
N.~Huang\textsuperscript{6} \and 
K.~D.~Irwin\textsuperscript{4,22,5} \and 
O.~B.~Jeong\textsuperscript{6} \and 
M.~Jonas\textsuperscript{1} \and 
A.~Jones\textsuperscript{21} \and 
T.~S.~Khaire\textsuperscript{13} \and 
A.~M.~Kofman\textsuperscript{18} \and 
M.~Korman\textsuperscript{17} \and 
D.~L.~Kubik\textsuperscript{1} \and 
S.~Kuhlmann\textsuperscript{7} \and 
C.-L.~Kuo\textsuperscript{4,22,5} \and 
A.~T.~Lee\textsuperscript{6,23} \and 
E.~M.~Leitch\textsuperscript{2,8} \and 
A.~E.~Lowitz\textsuperscript{2} \and 
S.~S.~Meyer\textsuperscript{2,9,10,8} \and 
D.~Michalik\textsuperscript{21} \and 
J.~Montgomery\textsuperscript{11} \and 
A.~Nadolski\textsuperscript{18} \and 
T.~Natoli\textsuperscript{24} \and 
H.~Nguyen\textsuperscript{1} \and 
G.~I.~Noble\textsuperscript{11} \and 
V.~Novosad\textsuperscript{13} \and 
S.~Padin\textsuperscript{2} \and 
Z.~Pan\textsuperscript{2,10} \and 
P.~Paschos\textsuperscript{9} \and 
J.~Pearson\textsuperscript{13} \and 
C.~M.~Posada\textsuperscript{13} \and 
W.~Quan\textsuperscript{2,10} \and 
A.~Rahlin\textsuperscript{1,2} \and 
D.~Riebel\textsuperscript{21} \and 
J.~E.~Ruhl\textsuperscript{17} \and 
J.T.~Sayre\textsuperscript{15} \and 
E.~Shirokoff\textsuperscript{2,8} \and 
G.~Smecher\textsuperscript{25} \and 
J.~A.~Sobrin\textsuperscript{2,10} \and 
A.~A.~Stark\textsuperscript{26} \and 
J.~Stephen\textsuperscript{9} \and 
K.~T.~Story\textsuperscript{4,22} \and 
A.~Suzuki\textsuperscript{23} \and 
K.~L.~Thompson\textsuperscript{4,22,5} \and 
C.~Tucker\textsuperscript{3} \and 
L.~R.~Vale\textsuperscript{12} \and 
K.~Vanderlinde\textsuperscript{24,27} \and 
J.~D.~Vieira\textsuperscript{18,28} \and 
G.~Wang\textsuperscript{7} \and 
N.~Whitehorn\textsuperscript{16} \and 
V.~Yefremenko\textsuperscript{7} \and 
K.~W.~Yoon\textsuperscript{4,22,5} \and 
M.~R.~Young\textsuperscript{27} \and 
}
\institute{
\textsuperscript{1}{Fermi National Accelerator Laboratory, MS209, P.O. Box 500, Batavia, IL, 60510, USA} \\
\textsuperscript{2}{Kavli Institute for Cosmological Physics, University of Chicago, 5640 South Ellis Avenue, Chicago, IL, 60637, USA} \\
\textsuperscript{3}{School of Physics and Astronomy, Cardiff University, Cardiff CF24 3YB, United Kingdom} \\
\textsuperscript{4}{Kavli Institute for Particle Astrophysics and Cosmology, Stanford University, 452 Lomita Mall, Stanford, CA, 94305, USA} \\
\textsuperscript{5}{SLAC National Accelerator Laboratory, 2575 Sand Hill Road, Menlo Park, CA, 94025, USA} \\
\textsuperscript{6}{Department of Physics, University of California, Berkeley, CA, 94720, USA} \\
\textsuperscript{7}{High-Energy Physics Division, Argonne National Laboratory, 9700 South Cass Avenue., Argonne, IL, 60439, USA} \\
\textsuperscript{8}{Department of Astronomy and Astrophysics, University of Chicago, 5640 South Ellis Avenue, Chicago, IL, 60637, USA} \\
\textsuperscript{9}{Enrico Fermi Institute, University of Chicago, 5640 South Ellis Avenue, Chicago, IL, 60637, USA} \\
\textsuperscript{10}{Department of Physics, University of Chicago, 5640 South Ellis Avenue, Chicago, IL, 60637, USA} \\
\textsuperscript{11}{Department of Physics and McGill Space Institute, McGill University, 3600 Rue University, Montreal, Quebec H3A 2T8, Canada} \\
\textsuperscript{12}{NIST Quantum Devices Group, 325 Broadway Mailcode 817.03, Boulder, CO, 80305, USA} \\
\textsuperscript{13}{Materials Sciences Division, Argonne National Laboratory, 9700 South Cass Avenue, Argonne, IL, 60439, USA} \\
\textsuperscript{14}{Canadian Institute for Advanced Research, CIFAR Program in Gravity and the Extreme Universe, Toronto, ON, M5G 1Z8, Canada} \\
\textsuperscript{15}{CASA, Department of Astrophysical and Planetary Sciences, University of Colorado, Boulder, CO, 80309, USA } \\
\textsuperscript{16}{Department of Physics and Astronomy, University of California, Los Angeles, CA, 90095, USA} \\
\textsuperscript{17}{Department of Physics, Center for Education and Research in Cosmology and Astrophysics, Case Western Reserve University, Cleveland, OH, 44106, USA} \\
\textsuperscript{18}{Department of Astronomy, University of Illinois at Urbana-Champaign, 1002 West Green Street, Urbana, IL, 61801, USA} \\
\textsuperscript{19}{Harvey Mudd College, 301 Platt Boulevard., Claremont, CA, 91711, USA} \\
\textsuperscript{20}{Department of Physics, University of Colorado, Boulder, CO, 80309, USA} \\
\textsuperscript{21}{University of Chicago, 5640 South Ellis Avenue, Chicago, IL, 60637, USA} \\
\textsuperscript{22}{Deptartment of Physics, Stanford University, 382 Via Pueblo Mall, Stanford, CA, 94305, USA} \\
\textsuperscript{23}{Physics Division, Lawrence Berkeley National Laboratory, Berkeley, CA, 94720, USA} \\
\textsuperscript{24}{Dunlap Institute for Astronomy \& Astrophysics, University of Toronto, 50 St. George Street, Toronto, ON, M5S 3H4, Canada} \\
\textsuperscript{25}{Three-Speed Logic, Inc., Vancouver, B.C., V6A 2J8, Canada} \\
\textsuperscript{26}{Harvard-Smithsonian Center for Astrophysics, 60 Garden Street, Cambridge, MA, 02138, USA} \\
\textsuperscript{27}{Department of Astronomy \& Astrophysics, University of Toronto, 50 St. George Street, Toronto, ON, M5S 3H4, Canada} \\
\textsuperscript{28}{Department of Physics, University of Illinois Urbana-Champaign, 1110 West Green Street, Urbana, IL, 61801, USA} \\
}


\maketitle

\begin{abstract}

SPT-3G is a polarization-sensitive receiver, installed on the South Pole Telescope, that measures the anisotropy of the cosmic microwave background (CMB) from degree to arcminute scales.
The receiver consists of ten 150~mm-diameter detector wafers, containing a total of $\sim 16,000$ transition-edge sensor (TES) bolometers observing at 95, 150, and 220~\si{\giga\hertz}.
During the 2018-2019 austral summer, one of these detector wafers was replaced by a new wafer fabricated with Al-Mn TESs instead of the Ti/Au design originally deployed for SPT-3G.
We present the results of in-lab characterization and on-sky performance of this Al-Mn wafer, including electrical and thermal properties, optical efficiency measurements, and noise-equivalent temperature.
In addition, we discuss and account for several calibration-related systematic errors that affect measurements made using frequency-domain multiplexing readout electronics.


\end{abstract}

\section{Introduction}
Observations of the cosmic microwave background (CMB) by telescopes with arcminute resolution are useful in a wide range of measurements impacting cosmology and particle physics.
On small angular scales, better measurements of the CMB power spectrum and gravitational lensing will improve constraints on the sum of the neutrino masses and light relic particles~\cite{Henning:2017nuy,Wu:2019hek,Sherwin:2016tyf}.
While maps of gravitational lensing can be used to remove lensing contamination from low-resolution CMB surveys targeting inflationary B-modes~\cite{Manzotti:2017net,Ade:2018gkx}, high-resolution surveys can also search for inflationary B-modes directly in their own data.
Finally, high-resolution CMB maps can be used to discover massive clusters of galaxies~\cite{Bleem:2014iim,Hilton:2017gal}, dusty star-forming galaxies~\cite{Vieira:2009ru,Gralla:2019jyc}, and new transient sources~\cite{Whitehorn:2016njg}.

SPT-3G is a new receiver for the 10-meter South Pole Telescope (SPT) that was deployed in 2017 targeting these science goals~\cite{Benson:2014qhw,Bender:2018dwl}.
Its focal plane consists of $\sim 16,000$ transition-edge sensor (TES) bolometers split between ten 150~mm-diameter silicon wafers.
The detectors that were deployed during the 2017 and 2018 observing seasons used a 4-layer titanium/gold TES design fabricated at Argonne National Laboratory~\cite{Carter:2018unr}; however, in parallel we also developed an aluminum-manganese TES design similar to the ones developed by NIST for SPTpol~\cite{Henning:2012fz} and Advanced ACTPol~\cite{Li:2016}.
During the 2018-2019 austral summer, we replaced one of the deployed detector wafers with an Al-Mn wafer that we had characterized extensively in the lab.
In this paper, we describe the fabrication and lab characterization of the Al-Mn detectors, as well as their on-sky performance, which is comparable to that of the Ti/Au detectors used in SPT-3G.

\section{Design and Fabrication}
TESs fabricated with Al-Mn alloys have several attractive features.
Since Al-Mn is an alloy, it can be deposited in single step without requiring multiple layers to achieve the target critical temperature ($T_c$), thus simplifying the fabrication process.
The $T_c$ and the normal resistance ($R_n$) of the devices can also be tuned relatively independently of each other.
The $R_n$ of the film can be tuned by adjusting the thickness and geometry, while the $T_c$ can be grossly tuned by adjusting the Mn doping and then finely adjusted by controlled heating of the film after deposition~\cite{Li:2016}.
These features make Al-Mn a promising TES material for future CMB experiments, such as CMB-S4~\cite{Abazajian:2016yjj}.

\begin{figure}
\begin{center}
\includegraphics[width=\linewidth]{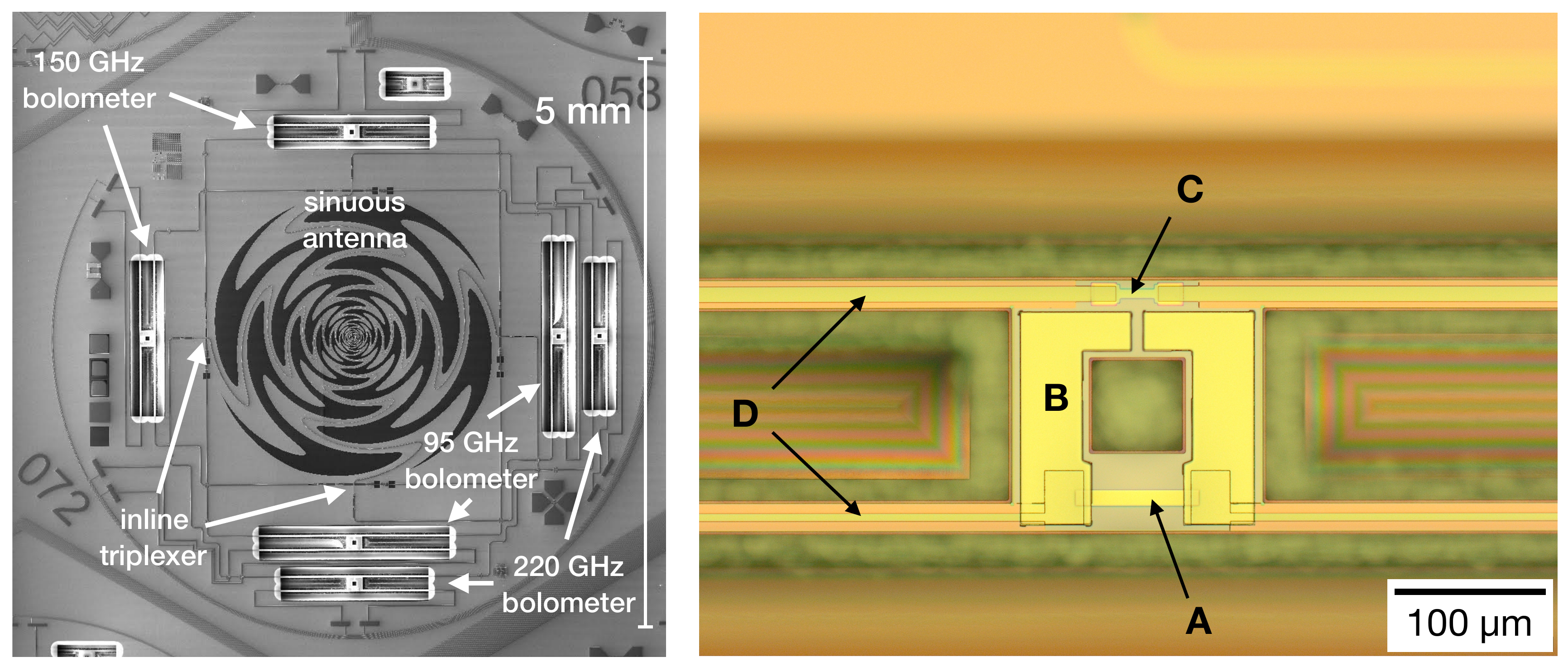}
\caption{
\emph{Left:} Layout of the trichroic SPT-3G pixel~\cite{Posada:2016}. Signals from the sky couple to a broadband dual-polarization sinuous antenna, are split into three observing bands by inline triplexers, and are detected by bolometers with Al-Mn TES films.
\emph{Right:} Bolometer island, showing (A) Al-Mn TES, (B) Pd heat capacity added to the island to stabilize the bolometer, (C) 20~\si{\ohm} load resistor for signals from the antenna, and (D) SiNx legs that mechanically support the suspended island and define the thermal conductivity of the bolometer.
(Color figure online)
\label{fig:fabphotos}
}
\end{center}
\end{figure}

Except for the features on the bolometer island, the fabrication of this Al-Mn detector wafer at Argonne was similar to previous SPT-3G detectors, as detailed in~\cite{Posada:2016}.
The layout of the SPT-3G pixel and the Al-Mn bolometer island are shown in Fig.~\ref{fig:fabphotos}.
The TES films are 75~\si{\nano\meter} thick, sputtered from aluminum doped with 850~ppm manganese by atomic percentage, capped with 15~\si{\nano\meter} Ti and 15~\si{\nano\meter} Au.
The 15~\si{\nano\meter} Au cap layer protects the TES film after deposition, and the 15~\si{\nano\meter} Ti layer prevents the formation of a compound between the Au cap and the Al in the TES film.
The TES film was patterned using lift-off to define 10~\si{\micro\meter} by 70~\si{\micro\meter} devices.
After TES film deposition, the wafer was baked at 180~C for 10~min to adjust the $T_c$ to the target of 420~\si{\milli\kelvin}.
The Au cap is a normal metal, so it reduces $T_c$ through the proximity effect, but the change in $T_c$ is small because the Au layer is thin.
A 600~\si{\nano\meter} palladium film was also deposited on the island, overlapping with the TES, to increase the total heat capacity and improve the electrothermal stability of the bolometer.
This Pd film was thinner than the 850~\si{\nano\meter} films used previously in SPT-3G, in order to reduce the thermal time constant of the detectors.

\section{Detector Performance}
\subsection{Electrothermal properties}\label{sec:electrothermal}
The yield of functional detectors on the Al-Mn detector wafer is very similar to the rest of SPT-3G~\cite{Bender:2018dwl,Bender:2019}.
Of 1572 wired detectors, 91.7\% have nominal resistances measured at 300~\si{\kelvin} and 89.9\% correspond to measured resonances in the readout electronics.
Losses are due to a combination of on-wafer opens and shorts, defects in wirebonding, and open channels in the multiplexing readout electronics ($<2$\%).
During routine observing during May 2019, an average of 79.4\% of all detectors were actually operated in the transition.
Nearly all of the additional losses (9.6\% of the 10.4\% difference) are due to a minimum requirement on optical responsivity.

\begin{figure}
\begin{center}
\includegraphics[width=0.8\linewidth]{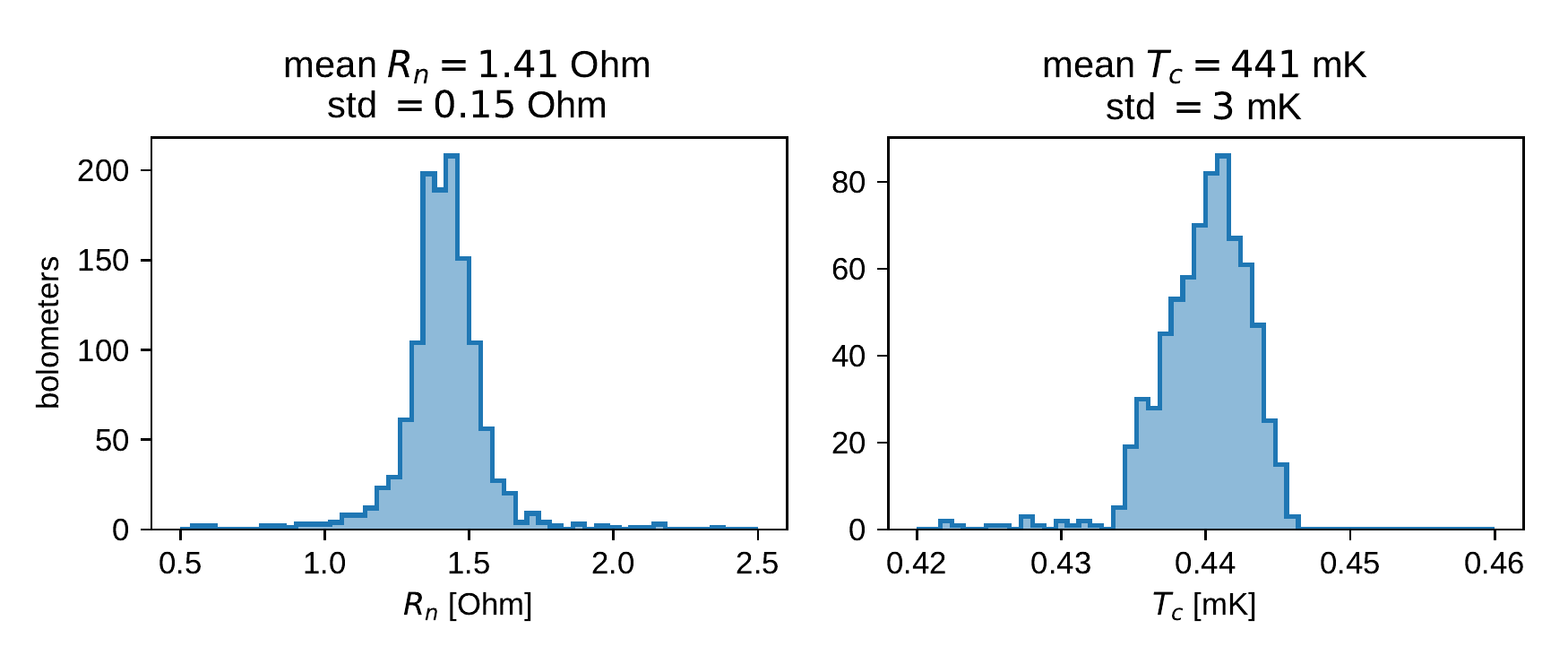}
\caption{
\emph{Left:} Distribution of normal resistances measured in the field on SPT-3G, with parasitic resistance subtracted. 
Target resistance was 1.7~\si{\ohm} compared with the measured mean of 1.41~\si{\ohm}.
\emph{Right:} Distribution of TES critical temperatures measured in lab tests prior to deployment.
The lab measurements shown here were performed on only one third of the wafer.
The target $T_c$ was 420~\si{\milli\kelvin} compared to the measured mean of 441~\si{\milli\kelvin}.
This difference in $T_c$ is typical of the batch-to-batch variation of devices fabricated at Argonne.
(Color figure online)
\label{fig:RnTc}
}
\end{center}
\end{figure}

The TES $R_n$ and $T_c$ were measured in the lab prior to deployment on approximately one third of the detectors on the wafer.
The measured $T_c$ of $441 \pm 3$~\si{\milli\kelvin} was acceptably close to the 420~\si{\milli\kelvin} target and had excellent uniformity of better than 1\% across the wafer, as shown in Fig.~\ref{fig:RnTc}.
This difference in $T_c$ is typical of the batch-to-batch variation of devices fabricated at Argonne.
The $R_n$ of each bolometer in the deployed system is estimated as
\begin{equation}
\hat{R}_n = \frac{R_{\textrm{total}}}{c(i)} - R_p(f_b),
\end{equation}
where $R_{\textrm{total}}$ is the total resistance measured by the readout system in the normal state, $R_p(f_b)$ is the average parasitic impedance as a function of bolometer bias frequency\footnote{The parasitic impedance measured in the superconducting state is dominated by reactances that produce a characteristic frequency-dependence across all detector wafers and readout modules.} $f_b$, and $c(i)$ is an empirical correction factor accounting for stray capacitance as a function of resonator index in the LC multiplexer chips as described in \cite{Dutcher:2018fod}.
An $R_n$ of $1.41 \pm 0.15$~\si{\ohm} was measured in the deployed wafer, also shown in Fig.~\ref{fig:RnTc}.
Of the variance in $R_n$, 0.06~\si{\ohm} is attributable to using a model for $R_p$ rather than bolometer-by-bolometer measurements, and much of the remainder is likely due to residual calibration uncertainties in the readout electronics; DC four-wire measurements of $R_n$ typically show much lower scatter than measurements made with our frequency-domain multiplexing (fMux) readout electronics.
The variance in $R_n$ in Fig.~\ref{fig:RnTc} should therefore not be interpreted as entirely due to physical variation in $R_n$ across the wafer.

\begin{figure}
\begin{center}
\includegraphics[width=\linewidth]{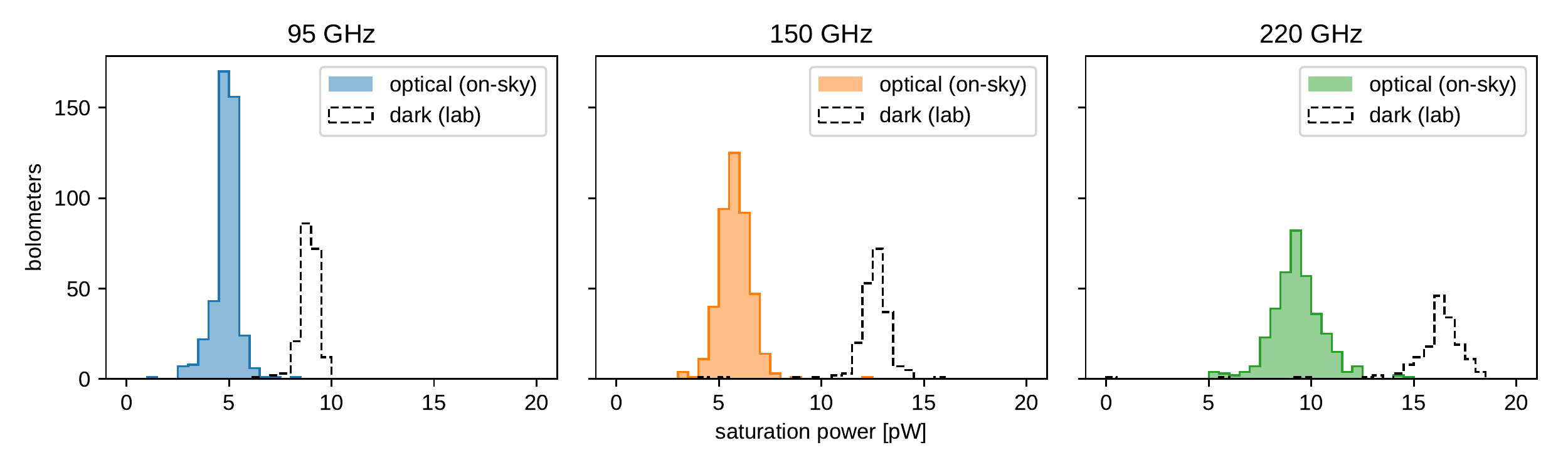}
\caption{
Saturation powers for 95~\si{\giga\hertz} (\emph{left}), 150~\si{\giga\hertz} (\emph{center}), and 220~\si{\giga\hertz} (\emph{right}) bolometers with and without an optical load from the sky, measured with a bath temperature of 315~\si{\milli\kelvin}.
Dark saturation powers \emph{(black dashed)} were measured in the lab prior to deployment.
Optical saturation powers \emph{(solid)} are the median power, bolometer-by-bolometer, measured over the month of June 2019 at an elevation of 63.5~deg (1.27~airmasses).
All power measurements are adjusted by assuming a 0.9~\si{\nano\henry} inductance in series with the shunt resistor in the readout (see text for details).
(Color figure online)
\label{fig:Psat}
}
\end{center}
\end{figure}

The saturation powers were measured both in the lab and with optical load in the field, shown in Fig.~\ref{fig:Psat}.
The detectors are operated under a voltage bias by placing a $30\si{\milli\ohm}$ shunt resistor in parallel with the TES.
Saturation powers are calculated assuming that there is an additional 0.9~\si{\nano\henry} stray inductance in series with the shunt resistor.
Although we do not have a complete model of all parasitic impedances in the readout circuit, we find that this value is similar to previous measurements of this stray inductance, and including this parasitic impedance alone is sufficient to eliminate the spurious dependence of the saturation power on bias frequency\footnote{Note that saturation powers reported in \cite{Dutcher:2018fod} do not include this correction.}.
Since the saturation power depends on the thermal conductivity of the SiNx legs supporting the bolometer island, its value should be independent of the frequency at which it is biased by the readout electronics.
Relative to the saturation powers measured using nominal values of components in the readout circuit, the saturation powers are increased by
\begin{equation}
\hat{P}_{\textrm{sat}} = \frac{\sqrt{R_{sh}^2 + (2 \pi f_b L_{sh})^2}}{R_{sh}} \frac{R_{\textrm{total}} - R_p}{R_{\textrm{total}}} P^{\textrm{nominal}}_{\textrm{sat}},
\end{equation}
where $P^{\textrm{nominal}}_{\textrm{sat}}$ the saturation power assuming the nominal value $R_{sh}$ of the shunt resistor, $f_b$ is the TES bias frequency, and $L_{sh}$ is the inductance in series with the shunt resistor.
The target saturation power for this wafer, as for previous SPT-3G detectors, is 10~/~15~/~20~\si{\pico\watt} for 95~/~150~/~220~\si{\giga\hertz}.
Operated at a base temperature of 280~\si{\milli\kelvin}, the median dark saturation powers were measured to be 10.6~/~14.8~/~19.1~\si{\pico\watt}, in excellent agreement with the target.
During 2019 we began operating the SPT-3G receiver at a higher base temperature of 315~\si{\milli\kelvin} to improve the $^3$He/$^4$He refrigerator cycle efficiency, so the saturation powers shown Fig.~\ref{fig:Psat} are smaller but still have sufficient margin to avoid saturation.

\subsection{Optical properties}
\begin{figure}
\begin{center}
\includegraphics[width=0.7\linewidth]{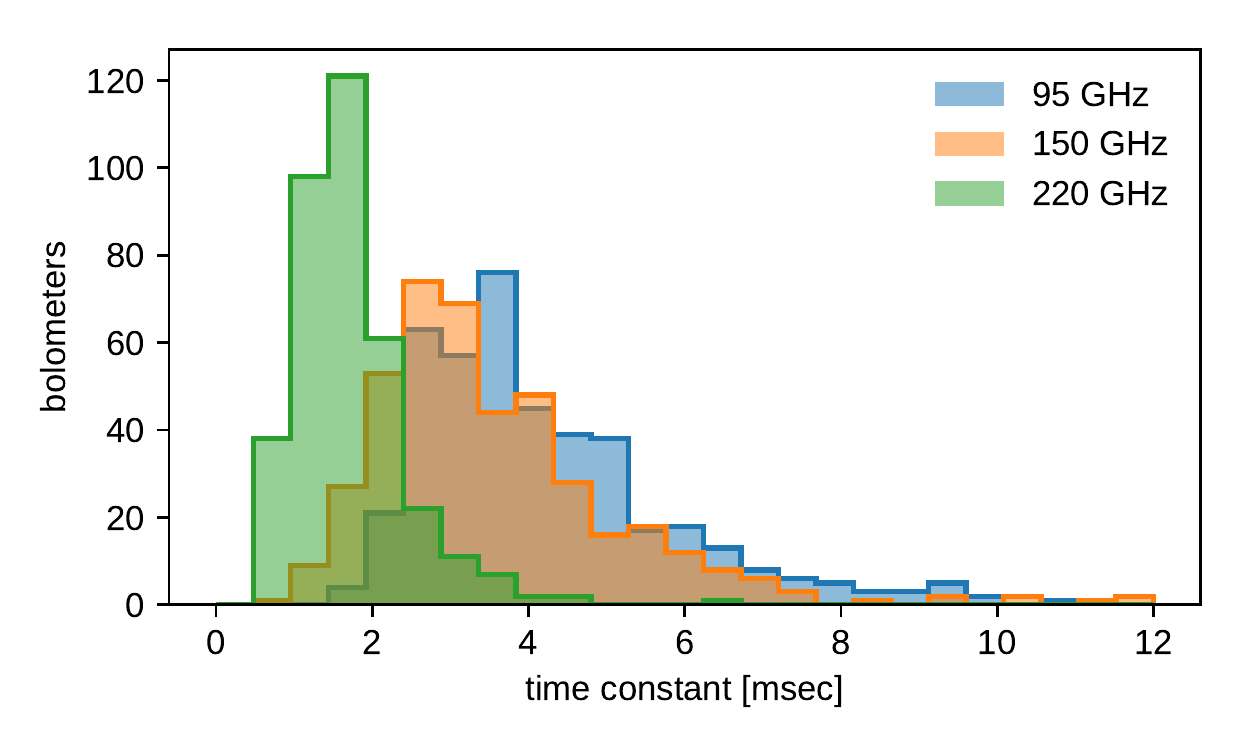}
\caption{
Optical time constant measured \emph{in situ} with a chopped thermal source behind the telescope secondary mirror.
Bolometer time constants are well above the threshold at which detectors become unstable but still fast enough to preserve good sensitivity to point sources.
(Color figure online)
\label{fig:tau}
}
\end{center}
\end{figure}

Optical time constants are measured regularly during observing using a chopped thermal source in the center of the SPT-3G secondary mirror.
The median time constants by bolometer are shown in Fig.~\ref{fig:tau} and average 4.4~/~3.6~/~1.7~\si{\milli\second} for 95~/~150~/~220~\si{\giga\hertz} detectors, when measured with atmospheric loading at an elevation of 67.5~deg.
The electrical power is highest for the 220~\si{\giga\hertz} detectors, which results in the smaller optical time constants for this band.
At the SPT-3G scan speed of 1~deg/sec on the azimuth bearing (less on the sky), the 3~\si{\decibel} frequencies of these TESs correspond to angular multipoles between 30,000 and 90,000 depending on observing elevation and band.
Since the 3~\si{\decibel} point of telescope beam at 150~\si{\giga\hertz} is $\ell \sim 8,000$, the detectors are fast enough to preserve good sensitivity to point sources.

\begin{figure}
\begin{center}
\includegraphics[width=\linewidth]{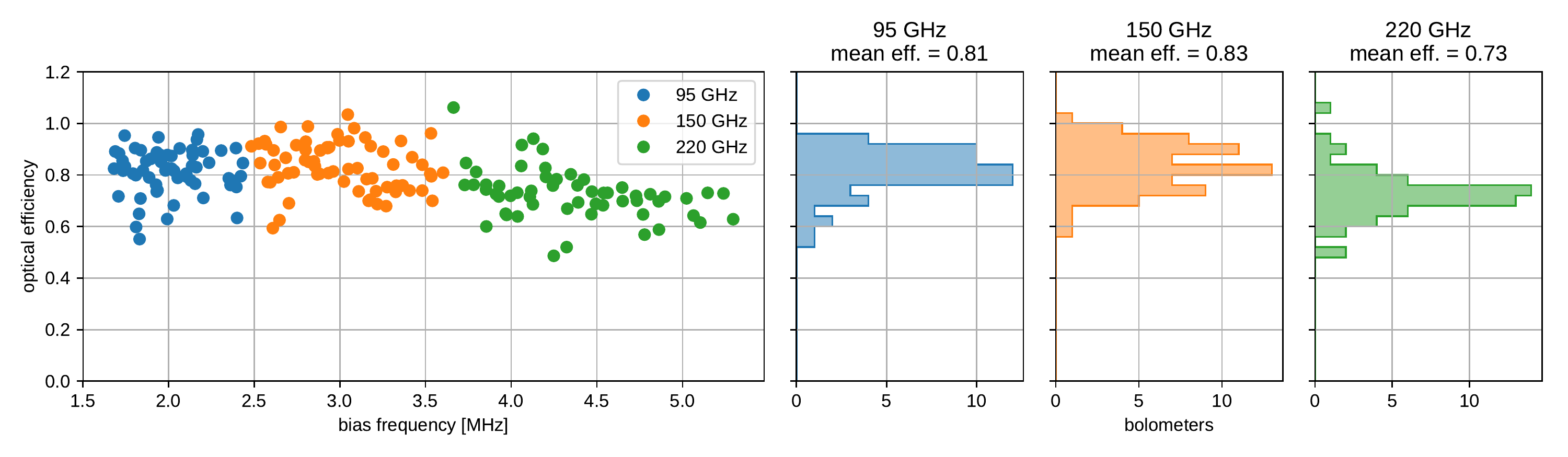}
\caption{
Optical efficiencies as a function of detector bias frequency, measured in a dark cryostat in the lab using a blackbody source whose temperature was varied between 8~\si{\kelvin} and 15~\si{\kelvin}.
This measurement includes the efficiency loss of alumina lenslets, the 3-layer PTFE anti-reflective coating on the lenslets~\cite{Nadolski:2018hck}, and all losses on the detector wafer from the antenna to the TES.
The relatively small number of detectors measured is because the cold load only illuminates 1/6 of the wafer.
(Color figure online)
\label{fig:opticaleff}
}
\end{center}
\end{figure}

To measure the optical efficiency, the detectors were illuminated by a blackbody source, and saturation powers were measured at eight temperatures of the blackbody between 8~\si{\kelvin} and 15~\si{\kelvin}. 
The measured saturation powers, $P_{\textrm{sat}}^{\textrm{nominal}}$, as a function of cold load temperature, $T_\mathrm{cl}$, are fit to the model
\begin{align}
\begin{split}\label{eqn:psatmodel}
P_{\textrm{sat}}^{\textrm{nominal}}(T_{cl}) = &P_0 + \left< C \right> T_{cl} \\
&- \eta \frac{R_{sh}}{\sqrt{R_{sh}^2 + (2\pi f_b L_{sh})^2}} \frac{R_{\textrm{total}}}{R_{\textrm{total}} - R_p}\\
&\times \int \frac{h \nu}{\exp{(h\nu / kT_{cl})} - 1} T_{\textrm{triplexer}}(\nu) T_{\textrm{filter}}(\nu) ~d\nu,
\end{split}
\end{align}
where the fit parameters $P_0$ and $\eta$ represent respectively the saturation power with no optical load and the optical efficiency.
Several additional factors are included to correct for systematic biases in the measurement:
\begin{itemize}
\item \emph{Wafer heating:} The cold load aperture illuminates only one triangular hextant of the wafer, through three low-pass metal-mesh filters (cutoff of 300~\si{\giga\hertz}), and other detectors are blocked by an aluminum plate maintained at the base temperature near 250~\si{\milli\kelvin}.
The saturation powers of blocked detectors show a small linear dependence on the cold load temperature, suggesting that the radiative load is heating up the wafer slightly.
We fit data from each blocked detector to a linear model of the form $P_{\textrm{sat}}^{\textrm{nominal}}(T_{cl}) = P_0 + CT_{cl}$, compute the average $\left< C \right>$ of the slopes across all bolometers in each observing band, and then assume that the illuminated bolometers experience the same thermal effects as the blocked detectors by including the term $\left< C \right> T_{cl}$ in Eq.~\ref{eqn:psatmodel}.
The net effect of this term is to decrease the estimated optical efficiency.
\item \emph{Filter transmission:} The blackbody illuminates optically active detectors through a stack of three low-pass metal-mesh filters.
While we do not have high-quality measurements of all filters over the full observing band, a measurement of one filter over the 150 and 220~\si{\giga\hertz} bands has an average transmission of 96\%. For simplicity, we therefore assume each filter has a fixed frequency-independent transmission of 96\%, represented by the factor $T_{filter}(\nu)$ in Eq.~\ref{eqn:psatmodel}.
\item \emph{Triplexer response:} $T_{triplexer}(\nu)$ represents the frequency response of the inline triplexer for each band, which is taken from Sonnet simulations that assume a dielectric thickness typical of deployed wafers.
\item \emph{Inductance in series with shunt resistor:} As discussed in Section~\ref{sec:electrothermal}, there is a wiring inductance $L_{sh} \sim 0.9~\si{\nano\henry}$ in series with the $R_{sh} = 30~\si{\milli\ohm}$ shunt resistor.
This gives rise to a difference between the physical power incident on the TES, which is accounted for by the first factor in front of the integral in Eq.~\ref{eqn:psatmodel}.
\item \emph{Parasitic resistance in series with TES:} As discussed in Section~\ref{sec:electrothermal}, there is an impedance in series with the TES, so that the voltage bias produced by the shunt resistor is divided between the series resistance and the TES. This effect is accounted for by the second factor in front of the integral in Eq.~\ref{eqn:psatmodel}.
\end{itemize}
Fig.~\ref{fig:opticaleff} shows the total optical efficiency as a function of bias frequency, estimated from the fit of Eq.~\ref{eqn:psatmodel}.
The mean efficiencies are 81\%~/~83\%~/~73\% for 95~/~150~/~220~\si{\giga\hertz}.
The correction for the bias inductance has removed most of the spurious bias-frequency-dependence of optical efficiency within each band.
Some slight dependence may remain in the 220~\si{\giga\hertz} detectors, which could explain the lower estimate efficiency in that band.

In addition to the lab measurements, we also routinely measure the noise-equivalent temperature (NET) of the detectors during observing.
While NET is not an equivalent metric to optical efficiency, poor optical efficiency will degrade the NET.
The median NET during June 2019 for this Al-Mn wafer was 675~/~523~/~1919~$\si{\micro\kelvin}\sqrt{\si{\second}}$ for 95~/~150~/~220~\si{\giga\hertz} detectors, versus 640~/~511~/~1815~$\si{\micro\kelvin}\sqrt{\si{\second}}$ for all other detectors.
The small difference in NET between the Al-Mn wafer and the other Ti/Au SPT-3G wafers is comparable to the variation among the Ti/Au wafers, suggesting that overall performance and optical efficiency of the Al-Mn is similar to the Ti/Au.

\section{Status and Outlook}
We have fabricated an array of Al-Mn TES bolometers for the SPT-3G experiment.
Extensive lab characterization of this array indicates that it meets our design specifications for normal resistance, saturation power, critical temperature, and optical efficiency, with high uniformity in all of these parameters.
It was installed in the SPT-3G receiver and has been used in the 2019 observing season, where on-sky yield and NET are very similar to the other detector wafers.
The excellent overall performance indicates that Al-Mn TES bolometers produced by Argonne would be suitable for a next-generation CMB experiment such as CMB-S4.

\begin{acknowledgements}
The South Pole Telescope is supported by the National Science Foundation (NSF) through grant PLR-1248097. Partial support is also provided by the NSF Physics Frontier Center grant PHY-1125897 to the Kavli Institute of Cosmological Physics at the University of Chicago, and the Kavli Foundation and the Gordon and Betty Moore Foundation grant GBMF 947. Work at Argonne National Laboratory, including Laboratory Directed Research and Development support and use of the Center for Nanoscale Materials, a U.S. Department of Energy, Office of Science (DOE-OS) user facility, was supported under Contract No. DE-AC02-06CH11357. We acknowledge R. Divan, L. Stan, C.S. Miller, and V. Kutepova for supporting our work in the Argonne Center for Nanoscale Materials. Work at Fermi National Accelerator Laboratory, a DOE-OS, HEP User Facility managed by the Fermi Research Alliance, LLC, was supported under Contract No. DE-AC02-07CH11359. NWH acknowledges support from NSF CAREER grant AST-0956135. The McGill authors acknowledge funding from the Natural Sciences and Engineering Research Council of Canada, Canadian Institute for Advanced Research, and the Fonds de recherche du Qubec Nature et technologies. JV acknowledges support from the Sloan Foundation.
\end{acknowledgements}

\pagebreak
\printbibliography

\end{document}